\documentclass[11.1pt,epsf]{article}
\usepackage[english]{babel} \usepackage{amsmath, amssymb, multicol, amsfonts, amsthm, amscd, epsfig, enumerate}
\usepackage[latin1]{inputenc} \usepackage[T1]{fontenc} \usepackage{graphicx} \usepackage{geometry}

\selectfont

\begin{document}

\author{
\emph{\textbf{Nazaria Solferino}}\\ \small{ Economics Department,
University of Rome "Tor Vergata"} \\ \emph{\textbf{Viviana Solferino}}\\
\small{ Mathematics and Computer Science Department, University of
Calabria} }

\title{The Corporate Social Responsibility is just a twist in a
\textrm{M\"obius Strip } } \date{} \maketitle

\begin{abstract}
In recent years  economics agents and systems have became more and
more interacting and juxtaposed, therefore the social sciences
need to rely on the studies of physical sciences to analyze this
complexity in the relationships. According to this point of view
we rely on the geometrical model of the M\"obius strip used in the
electromagnetism which analyzes the moves of the electrons that
produce energy. We use a similar model in a Corporate Social
Responsibility context to devise a new cost function in order to
take into account of three positive crossed effects on the
efficiency: i)cooperation among stakeholders in the same sector;
ii)cooperation among similar stakeholders in different sectors and
iii)the stakeholders' loyalty towards the company. By applying
this new cost function to a firm's decisional problem we find that
investing in Corporate Social Responsibility activities is ever
convenient depending on the number of sectors, the stakeholders'
sensitivity to these investments and the decay rate to alienation.
Our work suggests a new method of analysis which should be
developed not only at a theoretical but also at an empirical
level.
\end{abstract}



\maketitle \noindent {}
\medskip

\section{Introduction}
In recent years, in particular from the beginning of the 21st
century, the social sciences started to strongly rely on the
discoveries of physics of complexity to analyze complicated
relations between models and social phenomena (Urry, 2003). For
instance this is just the research field of the econophysics which
studies the applications of theories and methods developed by
Physics in order to solve problems in Economics (for more details
see Rosser, 2008). As in the studies of many physical systems,
also in the social sciences there is a growing attention to go
behind the traditional notions treating various agents as
separated and distinct essences (Urry, 2003; Giddens,1984).
Currently they are instead conceived as juxtaposed entities
related trough a nonlinear mechanism where causes and effects are
co-present and strongly integrated\footnote{" No party to a
relation is therefore a monadic or molar entity. Each is instead a
mutable function or the character of the mode-of-being related and
its capacity for
relationality"(Dillon, 2000)}.\\
\noindent In an even more globalized  world very complex
interactions characterize social and economic relationships.
Therefore we need models taking into account this complexity and
nonlinearity in the connections. Such links involve multiple
positive and negative feedback loops making systems interdependent
and interacting dissipatively with their environment.\\
\noindent In Economics this interdependence among systems and
among agents is just the core of the models of Corporate Social
Responsibility (since now on CSR), which consider the global
integration between firms and stakeholders, including workers,
customers and the full environment (see Becchetti et al., 2014).
The CSR implies a move from the maximization of the shareholders
wealth to the satisfaction of a more complex objective function in
which interests of the other stakeholders are taken into account.
On turn this creates  also benefits for the business. For instance
Becchetti et al.(2014) show that since more and more profit
maximizing firms are adopting CSR practices there must be
pecuniary benefits arising from them. The authors also document
that the CSR has the potential to generate several values
increasing effects by attracting better employees, enhancing their
intrinsec motivation and loyalty, reducing turnover rates,
improving the efficiency and by reducing operating costs. Moreover
Becchetti and al. (2015) show that the CSR firms which take into
account the workers well-being are less exposed to business risks
and profit volatility. Nevertheless CSR improves boosting sales
revenues, increases rivals costs and attracts more ethical
consumers, so that the firm can benefit from increases in her demand share.\\
\noindent All the above mentioned advantages can be seen as a sort
of \emph{ethical capital} accumulated trough the CSR practices,
which also requires the payment of additional costs. Becchetti et
al. (2014) underline, by using a dynamic model, the conditions
implying that such benefits overrun the costs. These advantages
can also be considered  as the result of the synergy which relates
each subsystem's and each agent's performance.\\
\noindent Thanks to this synergy net benefits from the
relationships across to the stakeholders by the virtue of their
connections to the firm and the net transactional benefits across
to the business system by the virtue of the  intra-organizational cooperation.\\
\noindent Therefore according to the CSR point of view  firms and
stakeholders can be depicted not as two distinct and unconnected
systems, but they are a cross-system where transfers occur in a
such a way that a business becomes a stakeholders' interest and
conversely stakeholders well-being becomes part of the business.
In this crossed-system  the output of each part is transferred
across them to become the others' input, so that these subsystems
are strongly overloaded and linked inextricably
together.\\
\noindent According to our point of view the best metaphor,
suggested by  the physical sciences, to approximate and represent
this new conceptualization of links in economics systems and
between agents is the \emph{M\"obius
strip}.\\
\noindent This is a topological enigma independently documented in
1858 by two mathematicians A. F. M\"obius and J.B. Listing. It is
a bend of paper given a $180$ degree twist prior to having its two
ends connected. The first use of the M\"obius strip as a metaphor
in the business relationships, on our knowledge, is that of Litz
(2008), who discusses an alternative approach to business family
and family business relationships.\\
\noindent In this work we aim to extend this approach to the CSR
analysis by extensively relying on the recently discoveries in the
electromagnetism. We assimilate firm and stakeholders' contributes
to the action of electrons travelling on a M\"obius strip which,
unlike a regular bend, return to a mirror reality in each count.
In particular we strictly follow the model of Yacubo et al.(2003)
who show that the electrons travelling on a M\"obius strip produce
energy of higher intensity or equivalently there is a lower energy
dissipation thanks to the decreased resistance by virtue of the
twist in the bend. We analyze how contributions of the economic
agents in a CSR context, thanks to the effect to the ethical
capital, produce higher benefits and a lower dissipation of the
costs thanks the augmented cooperation.\\
\noindent The paper is divided into four sections (including
introduction and conclusions). In the second section we describe
the building of the geometrical model for the electrons travelling
in a M\"obius strip. In the third section we investigate how to
apply this model to the behavior of firms and economics agents in
a CSR context. We define a new cost function that show the
convenience to invest in social responsible activities thanks to
three positive crossed effects on the efficiency: i)cooperation
among stakeholders in the same sector; ii)cooperation among
similar stakeholders in different sectors and iii)the
stakeholders' loyalty towards the company. We provide an example
of a firm's decisional problem which decides whether to invest in
social responsibility. Our analytical results show that this is
ever the optimal choice depending on the number of sectors, the
stakeholders' sensitivity to these investments and the decay rate
to alienation. In the fourth section we discuss our conclusions.
\section{How to build a geometrical model for the electrons
travelling in a M\"obius strip}
 The M\"obius strip is a bi-dimensional manifold
with only one face. It can be built from a strip of paper by
joining together its both ends after having twisted one of them a
half turn (see Figure 1).
\newpage
  \begin{figure}[ht]
     \begin{center}
     \includegraphics[width=8cm]{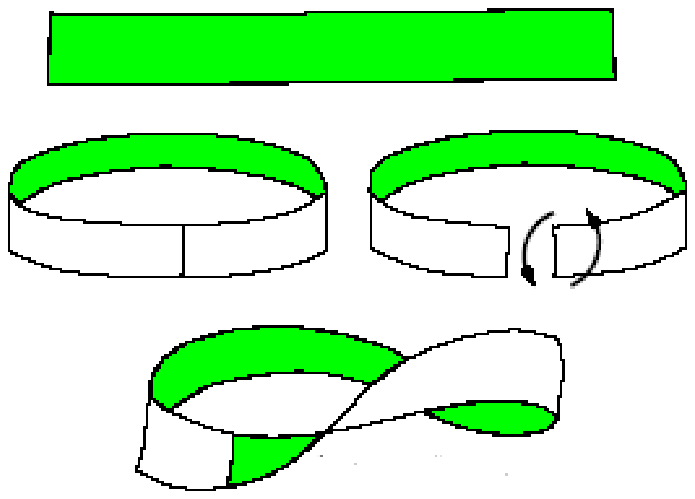}
     \end{center}
     \end{figure}
\small{\centerline{Figure 1: How to build a M\"obius strip}}
\vskip 1 pc \noindent The M\"obius strip has one side and a single
border and if we move along the centre line, the meridian, of the
strip we need to go through the circle twice in order to return to
the original position. This behavior is similar to that of the
electrons generating a flux periodicity of persistent currents in
a M\"obius strip in Yacubo et al. (2003), who describe it by using
the Hubbard model (1963). This last is the simplest model of
interacting particles (electrons) in a lattice and consists of a
Hamiltonian with only two terms: a \emph{kinetic term} which
represents the kinetic energy of electrons hopping between atoms
and a \emph{potential term} consisting of an on-site interaction
which represents the potential energy arising from the charges on
the electrons. If we assume that there are $N$ sites then we'll
say that if an electron tunnels from lattice site $j$ to site $l,$
its energy changes by an amount $-t_{jl}.$ This tunneling effect
is equivalent of annihilating the electron at site $j$ and
creating it again at site $l,$ so the portion of the Hamiltonian,
the kinetic term, dealing with tunneling can be written as
$$-\sum_{j,l=1}^N t_{jl}a^\dagger_{l}a_{j}$$
where $a^\dagger_{l},a_{j}$ are the fermion (since electrons are
fermions) creation and annihilation operators. For many practical
purposes it suffices to assume that $t_{jl}$ is none-zero, only
when $j$ and $l$ are the nearest neighbors in which case it is
usually approximated by a constant $t.$ Because of the electron
may tunnel also from lattice site $l$ to site $j,$ the Hamiltonian
becomes
$$-t\sum_{j,l=1}^N a^\dagger_{l}a_{j}+a^\dagger_{j}a_{l}$$
where $-t\sum_{j,l=1}^Na^\dagger_{j}a_{l}$ is defined Hermitian
conjugate and denoted by $h.c.$\\
\noindent The potential term is \vskip 0.5 pc
$$\sum_{k=1}^N \varepsilon_{k}a^\dagger_{k}a_{k}$$
\vskip 0.5 pc \noindent where $\varepsilon_{k}$ represents the
site energy and $a^\dagger_{k},a_{k}$ are the fermion creation and
annihilation operators at the site $k.$\\
\noindent Yacubo et al.(2003) consider electrons moving on a
M\"obius strip in the longitudinal directions on $2M$ wires and
transverse directions on $N$ wires. Specifically, starting from a
rectangular lattice including $N\times2M$ sites (see Figure 2),
the rectangle is then twisted by $180$ degrees and its two sides
are connected, such that longitudinal wire $1$ is attached to wire
$2M,$ wire $2$ is attached to wire $2M-1$ and so on (see Figure
3). The M\"obius strip so constructed includes $M$ longitudinal
wires with $2N$ sites on each one.\vskip 3 pc
\begin{figure}[ht]
     \begin{center}
     \includegraphics[width=9cm]{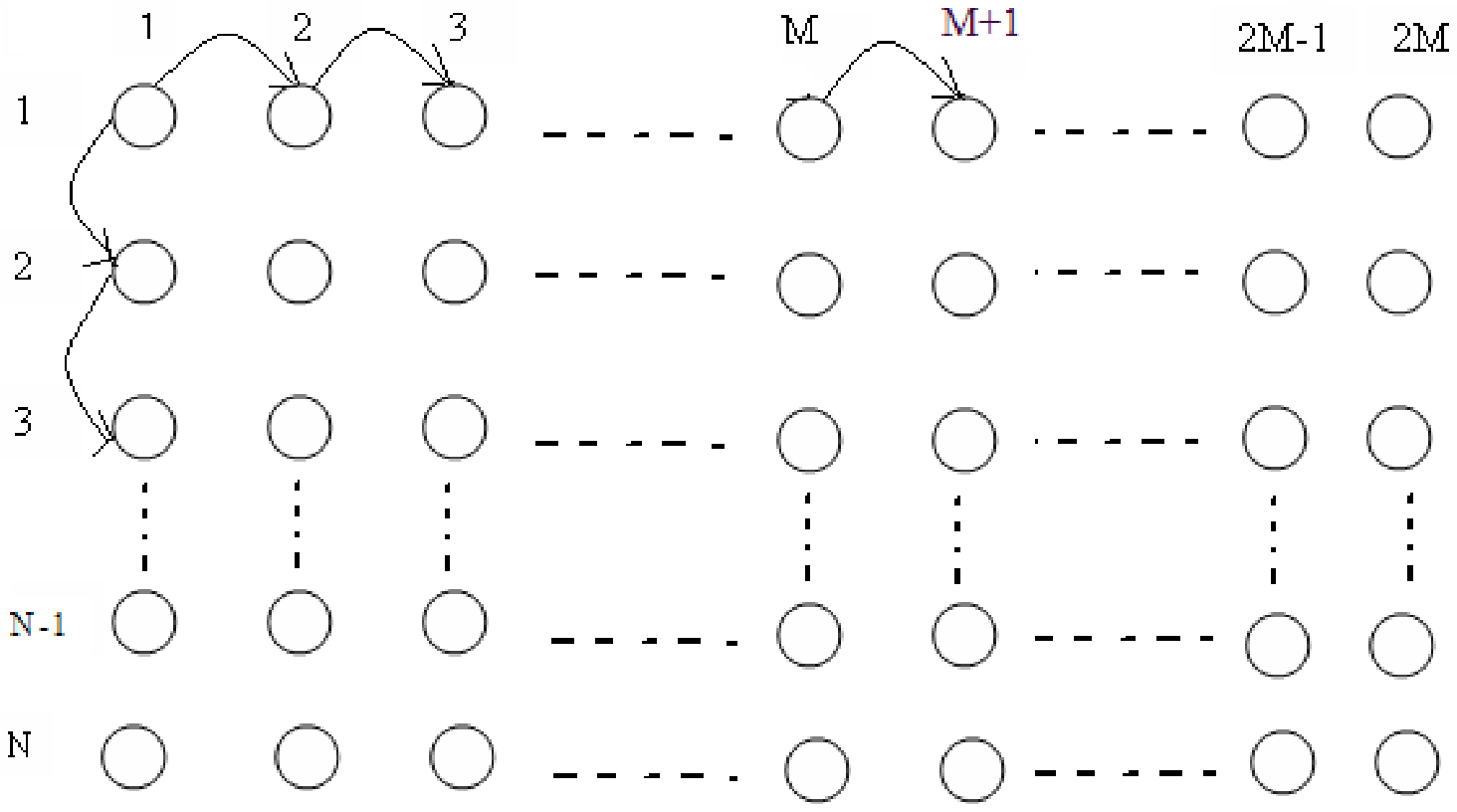}
     \end{center}
     \end{figure}
\small{\centerline{Figure 2: The electrons moving in a lattice
$N\times 2M.$}}
\newpage
\begin{figure}[ht]
     \begin{center}
     \includegraphics[width=9cm]{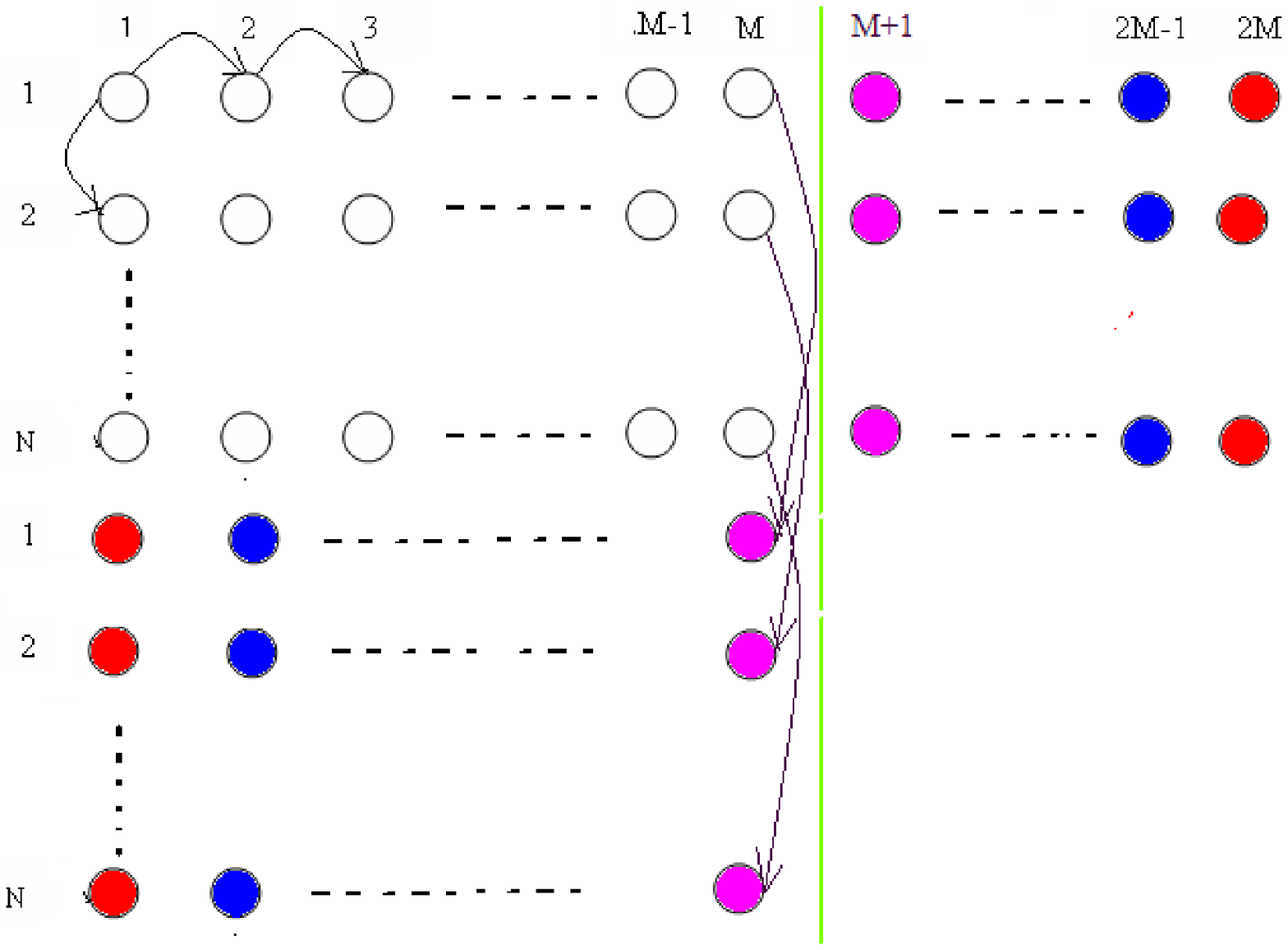}
     \end{center}
     \end{figure}
\small{\centerline{Figure 3: The electrons moving in a M\"obius
strip.The previous lattice has became a}} \small{\centerline{
lattice $2N\times M.$ The area behind the green line,after the
twist,shifted in the bottom} \small{\centerline{on the left.The
electrons in the column $M$ that tunneled in the $M+1$ column,now}
\small{\centerline{tunnel in the same column $M$ on the
corresponding replicated new element.}} \vskip 2 pc
 According to the Hubbard model (1963)
the Hamiltonian is then
\begin{equation}\label{eq1}
 H_{M\ddot{o}bius}=\sum_{n=1}^{2N}\sum_{m=1}^{M}[\varepsilon_{nm}a^\dagger_{nm}a_{nm}-t_1e^{-2i\pi\Phi/N}a^\dagger_{nm}a_{n+1m}]
\end{equation}
\vskip 0.5 pc
$$-t_2\sum_{n=1}^{2N}\sum_{m=1}^{M-1}a^\dagger_{nm+1}a_{nm}-\frac{t_2}{2}\sum_{n=1}^{2N}a^\dagger_{nM}a_{n+NM}+h.c.$$
\vskip 0.5 pc \noindent where $a_{nm}$ is the fermion operator at
the site $(n,m)$ with $n=1,2,...,2N$ and $m=1,2,...,M).$\\
\noindent The quantity $\varepsilon_{nm}$ is the site energy so
that \vskip 0.5 pc
$$\sum_{n=1}^{2N}\sum_{m=1}^{M}\varepsilon_{nm}a^\dagger_{nm}a_{nm}$$
\vskip 0.5 pc \noindent represents the potential term.\\
\noindent The kinetic term is made up of three  parts:
\begin{enumerate}
    \item
    $-t_1\sum\limits_{n=1}^{2N}\sum\limits_{m=1}^{M}e^{-2i\pi\Phi/N}a^\dagger_{nm}a_{n+1m}$
    measures the longitudinal hopping, where $e^{-2i\pi\Phi/N}$ measures the effect of the magnetic field
    accumulated along the longitudinal direction on each link and $t_1$ is the longitudinal  hopping amplitude;\\
    \item
    $-t_2\sum\limits_{n=1}^{2N}\sum\limits_{m=1}^{M-1}a^\dagger_{nm}a_{nm}$  measures the transverse
    hopping on $M-1$ longitudinal wires and $t_2$ is the transverse hopping amplitude;
    \item the transverse hopping on the last wire $M$ is measured
    by
    $-\frac{t_2}{2}\sum\limits_{n=1}^{2N}a^\dagger_{nM}a_{n+NM}.$
    Without the twist the electron would tunnel from the site $(n,M)$ to the site $(n,M+1).$ But, because of the twist, now the wire
    $M+1$ is attached to the wire $M$ becoming the same longitudinal wire with $2N$ sites on
    it. Therefore the site $(n,M+1)$ is now the site $(n+N,M)$ (see Figure 3).Obviously the sum is divided by two because the electrons
    tunnel only from (towards) the original $N$ sites.
\end{enumerate}
\section{The Economics of the CSR-M\"obius strip}
\subsection{How to build a CSR-M\"obius strip economics model}
In this section we aim to investigate whether what we have seen in
the previous one can be applied to firms and economics agents in a
CSR context. Are there some similarities between their activities
and contributions to production and the move of electrons in the
strip that produces energy? At a first sight we notice that
$-H_{M\ddot{o}bius}$ strongly approaches a benefits-costs
function. In fact, the energy dissipation measured by
$\varepsilon$ can be assimilated to the production costs
unrecovered trough the sell of the added value of the final
consumption good.
\\
\noindent Similarly, the terms with $t_1$ and  $t_2$ may represent
the benefits associated to the joint contributions of $N$
stakeholders or type of stakeholeders operating in $M$ sectors.\\
\noindent For instance in the generalized Leontief production
function analyzed in Diewert (1971) the interindustrial relations
of an economy are conventionally represented by a matrix in which
each column lists the monetary value of an industry's inputs and
each row lists the value of the industry's outputs. Each cell of
this matrix might correspond to the site $(n,m)$ of the electrons
in the strip (for instance see Iyetomi et al. 2010).\\
\noindent Nevertheless we think that in a context of CSR this
function does not take into account all the crossed effects that
social responsible activities can generate in terms of
productivity and costs saving (see Becchetti et al. 2014). In
particular some of these effects concern the externalities due to
the CSR benefits on the stakeholders, which on turn are
transferred into positive returns on the firm's traditional
activities.\noindent According to this point of view, we consider
a SR company with $n=1,2,..,N$ stakeholders or cluster of
stakeholders and $m=1,2,...,2M$ activities, where $m=1,2,...,M$
represents the traditional sectors of production of intermediate
goods, necessary to produce the final good $M,$ while
$m=M+1,...,2M$ are the specific activities devoted to the CSR. We
denote by $0\leq a_{nm}<1$ the contribution  of the stakeholder
$n$ in the sector $m$ measured as percentage per unit of a
product. For instance if $a_{11}=\frac{1}{5}$ we say the
stakeholder $1$ is able to produce the $20$ per cent of a unit in
a working hour. \noindent Like in a M\"obius strip also in a
social responsible firm the effects of a twist may be considered
as the returns due to the CSR activities on the stakeholders and
firm production, which therefore amplify the crossed contributions
of different stakeholders also operating in different sectors of
the company (see Figure 4). \vskip 1 pc
\begin{figure}[ht]
     \begin{center}
     \includegraphics[width=11cm]{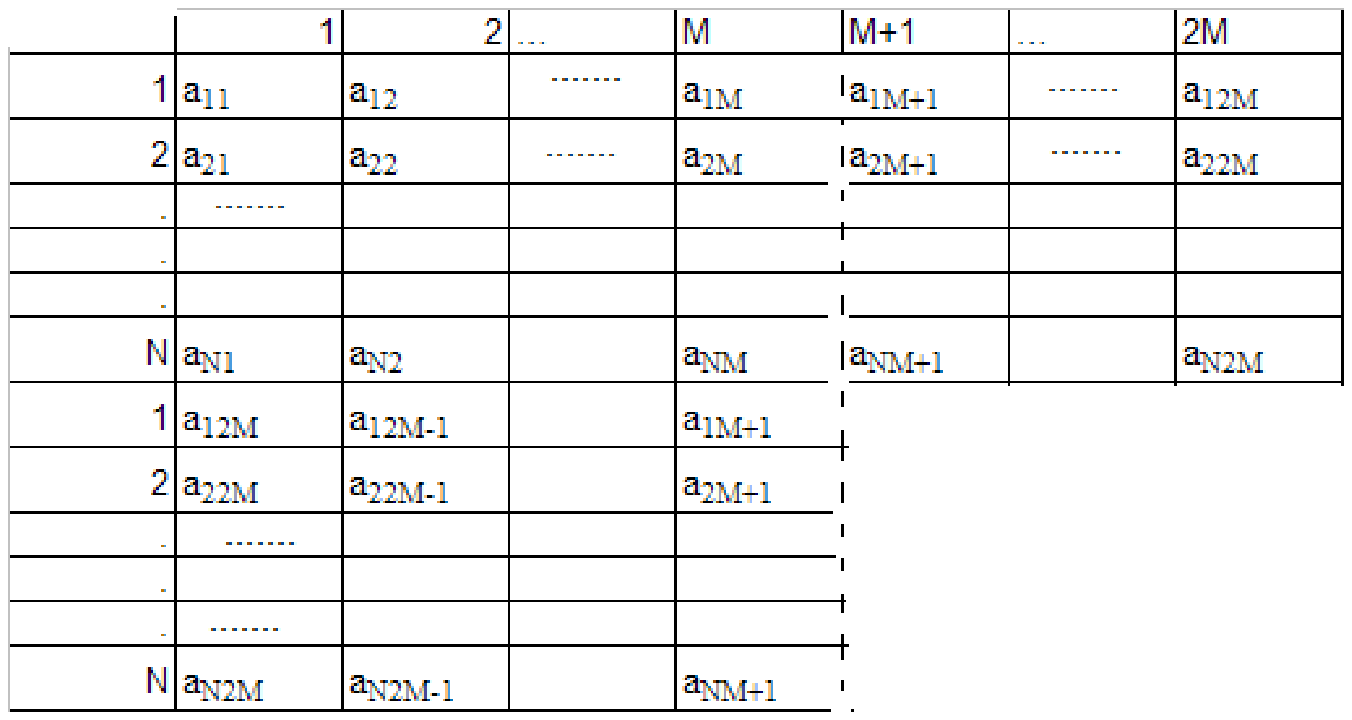}
     \end{center}
     \end{figure}
\small{\centerline{Figure 4: The matrix of
stakeholders'contributions in a CSR context.}}\vskip 1 pc
\noindent The stakeolder $1$ contributes with $a_{11}$ to the
production of the sector $1$ and with $a_{12}$ to the production
of the sector $2$ and so on. The stakeolder $2$ contributes with
$a_{21}$ to the production of the sector $1$ and with $a_{22}$ to
the production of the sector $2$ and so on. The same for all the
other stakeholders. The value of $a_{12M}$ measures the expected
additional contribution that the stakeholders $1$ would give
thanks to the social responsible activity $2M.$ The same for the
other social responsible activities which are ordered in such a
way that $2M$ is more relevant for the sector $1$, $2M-1$ is more
relevant for the sector $2,$ etc (for instance $2M$ could be seen
as  the social responsible activities dedicated to assure safety
work condition in sector $1,$ $2M-1$ those to assure safety work
condition in sector $2$ and so on). Therefore in this work we
propose the use of a new cost function for CSR companies suggested
by (\ref{eq1}), that in our case becomes: \vskip 0.5 pc
\begin{equation}\label{eq2}
 H_{CSR}=-\sum_{n=1}^{2N}\sum_{m=1}^{M}[c_{nm}-t_1(1-\delta)a_{nm}a_{n+1m}]+
 t_2\sum_{n=1}^{2N}\sum_{m=1}^{M-1}a_{nm+1}a_{nm}+\frac{t_2}{2}\sum_{n=1}^{2N}a_{nM}a_{n+NM}
\end{equation}
\vskip 0.5 pc \noindent where \vskip 0.5 pc
\begin{enumerate}
    \item $-\sum\limits_{n=1}^{2N}\sum\limits_{m=1}^{M}c_{nm}$ represents the sum
    of the costs supported by a company for social responsible activities
    devoted to each $n$ in the sector $m.$ The company
    can decide to give a prize also for the stakeholder's social
    responsible engagement and his  increased productivity in the traditional sectors, so that the
    cost can be different from zero for the $n=N+1,..., 2N$
    replicated stakeholders.\\
    \item
    $t_1\sum\limits_{n=1}^{2N}\sum\limits_{m=1}^{M}(1-\delta)a_{nm}a_{n+1m},$
    that we call \emph{the neighbouroud efficiency term}, measures the gains associated to the crossed contributions of $n$ in the sector
    $m$ with the nearest $n+1$ in the same sector. For instance if $a_{11}=\frac{1}{5}$ and
    $a_{21}=\frac{1}{7},$ when the SR stakeholder $1$ supports the
    stakeholder $2$ helping him to produce his share $\frac{1}{7},$ the stakeholder $1$ contributes with his ability of $\frac{1}{5}$
    to the production of $1+\frac{1}{7}$ units of the good.
    Therefore his total contribution is now $\frac{1}{5}\left(1+\frac{1}{7}\right).$
    Obviously also the stakeholder $2$ can support the stakeholder $1$ and this would correspond to Hermitian conjugate of this term.
    In the rest of the paper, to avoid excessive complexity, we don't consider the hermitian conjugate of (\ref{eq2}) because this doesn't affect our
    analysis. Moreover we assume that $0<\delta< 1$ is the decay rate due to the possible effect of
    alienation (caused for instance by satiety, low free time, etc.).
    Finally $t_1$ represents the sensitivity of the stakeholders' contributions to the SR activities devoted to
    them;\\
    \item
    $t_2\sum\limits_{n=1}^{2N}\sum\limits_{m=1}^{M-1}a_{nm+1}a_{nm},$ that we call \emph{sector cooperation efficiency term},
    measures the gains associated to the crossed contributions of  $n$ in the sector
    $m$ with the others type $n$ in the nearest sector $m+1$. Moreover $t_2$ (which can be equal or different from $t_1$) measures the sensitivity of the stakeholders
    contributions to the SR activities devoted to their and to other nearest sector.\\
    \item
    $\frac{t_2}{2}\sum\limits_{n=1}^{2N}a_{nM}a_{n+NM},$ that we call \emph{loyalty efficiency term},
    measures the gains associated to the increased productivity of each $n$ which contributes to the production of the final good
    $M$ twice: directly trough his own task and indirectly trough the
    increased efficiency and cooperative attitudes.\\
\end{enumerate}
Clearly all the above mentioned crossed effects could run among
more distant stakeholders and sectors. Nevertheless it is
reasonable to assume that this would imply not negligeable
transaction costs, necessary to raise useful and continuous
connections among them. Moreover the associated benefits should be
netted from the intermediate effects running among the nearest
ones. Therefore, all this things considered, it is possible to
assume, in our model, that those effects are very low and less
important for the company when she decides her investment
in CSR. \\
\noindent Moreover, we think that the main point is that SR firms
make specific investments (the sectors from $M+1$ to $2M$) to
foster stakeholders' socially responsible contributions and
productivity(which for examples are empirically measured by some
index as in the KLD metrics,see Becchetti et al. 2015) so to
reverse the upper  side of our matrix in the lower bound on the
left just as if we have two replicated stakeholders. The
traditional one making is own task, and the second is a sort of
replicated socially responsible stakeholders adding new
contributions to the firm. Therefore the order matters as
investments and return are specific into the firm. Obviously we
can imagine there are also externalities requiring no specific
orders, but they are difficult to measure and not related to
specific company's activities and investments while CSR measures
are specific for sectors and stakeholders so implying specific
returns. In particular the three above mentioned effects depend on
the extremely strick and precise conditions of how CSR investments
operate so that the twist is just a Mobius strip twist rather than
some less well-ordered reshuffling of cross-cutting effects across
the stakeholders.\\
\noindent In that follows  we aim to apply this function to a
general decisional  problem of a company which wants to minimize
the costs taking into account these crossed benefits due to the SR
activities.

\subsection{An application to a firm decisional problem with constant contributions and costs}
In this section we consider only one type of stakeholders and
specifically we assume that there are $N$ workers in
 $m=1,2,...,M$ traditional
sectors. We assume that the total production is equal to the sum
of the contributions of these workers, which could be measured in
term of pieces produced by worker in that sector in a working
hour, which is constant for each worker and sector, $a_{nm}=a,$
with $a\in \mathbb{R}$ and $0\leq a<1$ for all $n=1,2,...,N$ and
$m=1,2,...,M.$ Therefore if we denote by $p$ the price of the
final good and by $w$ the wages paid to workers, the firm's profit
function is: \vskip 0.5 pc
$$\pi=\sum_{n=1}^{N}\sum_{m=1}^{M}(p-w)a_{nm}=NMa(p-w).$$
\vskip 0.5 pc \noindent  We also assume that the company finances
the social responsible activities with an expense $c\geq 0$ equal
for each sector and worker and proportional to their
contributions, that is $c_{nm}=ca$ for all $n=1,2,...,N$ and
$m=1,2,...,M.$ Notice that this assumptions constant expense $c$
is not trivial and unrealistic. In fact, if we consider the same
type of stakeholders, in order to avoid any discrimination the
firm should invest, for each them, the same amount which is
proportional only to the own  contribution (meritocracy).
Otherwise it might have counterproductive effects (like envy,
frustration due to inequality, etc)instead of stimulating
cooperation and efficiency. In addition we suppose that the
worker's sensitivities $t_1$ and $t_2$ are equal and are related
to the investment in CSR through the function \vskip 0.5 pc
$$t_1=t_2=k{(ca)}^{\beta}$$
\vskip 0.5 pc \noindent where $k$ is a positive constant and
$\beta \in \mathbb{R}.$
\\
\noindent Under these assumptions, the company, for given values
$p$ and $w,$ wants to maximize the benefits associated to the
investment in CSR measured by the function (\ref{eq2}) that in
this case is \vskip 0.5 pc
\begin{equation}\label{eq3}
H_{CSR}(c)=-\sum_{n=1}^{2N}\sum_{m=1}^{M}[ca-t_1(1-\delta)a^2]+
 t_2\sum_{n=1}^{2N}\sum_{m=1}^{M-1}a^2+\frac{t_2}{2}\sum_{n=1}^{2N}a^2
\end{equation}
\vskip 0.5 pc \noindent subjected to \vskip 0.5 pc
\begin{equation}\label{eq4}
NMa[(p-w)-c]\geq 0
\end{equation}
\vskip 0.5 pc \noindent
 Obviously the constraint (\ref{eq4}) implies
that the firm can't expend in CSR more than what she would earn
without social responsible activities.\\
\noindent Simplifying (\ref{eq3}) we get \vskip 0.5 pc
\begin{equation}\label{eq5}
H_{CSR}(c)=-ca2NM+2kc^{\beta}NM(1-\delta)a^{2+\beta}+2kc^{\beta}N(M-1)a^{2+\beta}+kc^{\beta}Na^{4+\beta}
\end{equation}
\vskip 0.5 pc \noindent Therefore the company chooses the value of
$c$ that solves \vskip 0.5 pc
$$\frac{dH_{CSR}}{dc}=0$$
\vskip 0.5 pc \noindent under (\ref{eq4}), that is \vskip 0.5 pc
$$\frac{dH_{CSR}}{dc}=-a2NM+2\beta kc^{\beta-1}NM(1-\delta)a^{2+\beta}+2\beta kc^{\beta-1}N(M-1)a^{2+\beta}+k\beta c^{\beta-1}Na^{4+\beta}=0$$
\vskip 0.5 pc
$$c^{\beta-1}\beta k [2M(1-\delta)a^{1+\beta}+2
(M-1)a^{1+\beta}+ a^{3+\beta}]=2M.$$ \vskip 0.5 pc \noindent We
can distinguish three cases: \vskip 0.5 pc
\begin{itemize}
    \item [i)] for $\beta>1$
    $$c_1^*=\sqrt[\beta-1]{\frac{2M}{\beta k a^{1+\beta}[2M(2-\delta)-2
+ a^{2}]}}$$\vskip 0.5 pc which is a feasible solution only if
$c_1^*<p-w$. We can see that $c_1^*$ increases for high values of
$\delta$. In fact, being convenient to enforce workers'
sensitivity to SR to earn the high benefits due to $\beta>1$, the
company should invest more $c$ to counteract the negative effect
of $\delta$. Instead the optimal $c$ decreases for high values of
$\beta$ because no huge investments are necessary to stimulate
workers' sensitivity and the firm can save costs getting the same
great benefits. Finally, given the budget constraint, if there are
many sectors $M$ the company must invest a little amount $c$ for
each of them,
therefore $c$ decreases for high values of $M.$\\
    \item [ii)] for $\beta<1$
    $$c_2^*=\sqrt[1-\beta]{\frac{\beta k a^{1+\beta}[2M(2-\delta)-2
+ a^{2}]}{2M}}.$$ \vskip 0.5 pc Obviously the above mentioned
effects of $\delta, \beta$ and $M$ on the optimal value of $c$ are
reversed
when the workers are low sensitive to SR activities.\\
    \item [iii)] for $\beta=1$
    $$\frac{dH_{CSR}}{dc}= k a^2[2M(2-\delta)-2
+ a^2]-2M$$ which is constant. Therefore, if $$k
a^2[2M(2-\delta)-2 + a^2]-2M>0$$ it is ever convenient to invest
in CSR and the company chooses the optimal value of $c$ satisfying
\ref{eq4}, as she can easily recover the costs from the
proportional increase in $t$ for $k\geq 1.$ This condition is more
probably satisfied for high values of $k$ and $a$.
\end{itemize}

\section{Conclusions}
In the ongoing times characterized by an even more globalized
world, the reduction of distances thank to technologies make
people and systems (economic, social, cultural, etc) strongly
interrelated and juxtaposed. Therefore what happens somewhere
influences things happening elsewhere. From a theoretical point of
view to study these more interacting systems the traditional
economic models are improved also relying on the discoveries of
the physical sciences to take into account the several crossed
effects among the agents' actions. In particular in a CSR context
her related activities generate a sort of interlinked effects
which should be adequately analyzed. In this work we extensively
draw from the physical science and specifically from the
geometrical model of the M\"obius strip where the electrons
move in several directions to produce energy.\\
Similarly in a CSR context the social responsible activities have
the effects going in several directions which can increase the
stakeholders' productivity and efficiency so reducing production
costs. Therefore we devise a new cost-function where three crossed
effects are at work:$1)$ increases in the efficiency in virtue of
the augmented cooperation among the nearest stakeholders in the
same sector; $2)$ increases in efficiency in virtue of the
augmented cooperation among stakeholders in the nearest
sectors;$3)$ increases in the efficiency due to the augmented
stakeholders loyalty towards the vision of the company (and also
the management and the
shareholders)and so towards her final production.\\
\noindent We show how the benefits of the CSR in terms of those
three effects incentive the investment in CSR activities and we
also provide an example on how this new cost-function can be used
to analyze a simple SR firm's decisional problem. Our results show
that investing in CSR activities can ever be convenient depending
on the number of sectors, the stakeholders' sensitivity to these
investments and the decay rate to alienation.
\\We think that this approach could make light on effects in
productivity which  not have been adequately taken into account
and need to be more analyzed both at a theoretical and empirical
level. In particular proceeding from our theoretical model new
empirical measures on these crossed effects should be produced to
translate our model into reality.

\end{document}